# Laser-launched evanescent surface plasmon polariton field utilized as a direct coherent pumping source to generate emitted nonlinear four-wave mixing radiation

**Qun Zhang,**[1,2,*] **Ke Lin,**[1] **and Yi Luo**[1,2,3]

[1]*Hefei National Laboratory of Physical Sciences at the Microscale, University of Science and Technology of China, Hefei, Anhui 230026, P. R. China*
[2]*Department of Chemical Physics, University of Science and Technology of China, Hefei, Anhui 230026, P. R. China*
[3]*Department of Theoretical Chemistry, School of Biotechnology, Royal Institute of Technology, AlbaNova, S-106 91 Stockholm, Sweden*
*\*qunzh@ustc.edu.cn*

**Abstract:** We develop a concept of surface plasmon polaritons (SPPs) based four-wave mixing (4WM), in which a laser-launched evanescent SPP field is utilized as a coherent pumping source to involve directly in a nonlinear 4WM process at the dielectric/metal interface. Conversion efficiency of the resulting 4WM radiation is expected to be dramatically increased due to the local-field enhancement effect. Feasibility of implementing this concept at the air/gold film and graphene flake/gold film interfaces is further examined by numerical simulations. The concept shows intriguing promise for applications in newly emerging nanophotonics, optoelectronics, and active plasmonics.

- 2 -

## 1. Introduction

Surface plasmon polaritons (SPPs) are evanescent electromagnetic waves that propagate along the interface of a metal and a dielectric material as a result of collective electron oscillations coupled to an external optical field [1,2]. Triggered by rapid advances in nanofabrication and near-field optical microscopy, intensified SPPs-related research activities emerged in recent years [3-7]. Noticeably, the coherent nature of the evanescent SPP surface wave [1,2,8] has recently been "visualized" in a straightforward fashion in an elegant "double-slit" experiment with SPPs [9]. It is thus expected that SPPs may, in principle, be able to participate directly (and coherently) in conventional nonlinear optical processes of fundamental and practical interest, such as second-harmonic generation (SHG) [10-13], coherent anti-Stokes Raman scattering (CARS) [14], and four-wave mixing (4WM) [15-21].

In this contribution we put forward a concept pertinent to surface 4WM, the essence of which lies in that an evanescent SPP field localized at a dielectric/metal interface, though *nonradiative* (yet *coherent*), can be employed to interact *directly* with an additional propagating optical field, giving rise to an emitted nonlinear 4WM radiation in a coherent manner. We show that the conversion efficiency of the resulting 4WM emission can be significantly increased due to the local-field enhancement effect. We examine the concept by performing numerical simulations for two schemes of SPPs-based 4WM at the air/gold film interface. Furthermore, an extension to the graphene flake/gold film interface is presented. Our simulated two-dimensional wavelength-dependent angular distributions (2D-WAD) spectra (as well as the 4WM intensity distributions) provide straightforward guiding maps for further experimental verification.

To the best of our knowledge, the realization of nonlinear 4WM on metal surfaces by itself is not totally new. Very recently Novotny and co-workers demonstrated that nonlinear excitation of SPPs can be directly accomplished on gold surfaces by means of conventional optical 4WM, and further, both propagating and evanescent 4WM waves can be generated by vectorially adding the in-plane momenta of incident photons at the air/gold interface [17-19]. In contrast to these pioneering contributions in which the 4WM pumping sources are all propagating optical fields, the concept developed here distinguishes itself from them: An in-plane SPP wave, at the very moment of being established at a dielectric/metal interface, is employed to take the place of one of the optical pumping fields used for converting into a 4WM field; in other words, ensured by its coherent nature [8,9], the evanescent SPP field acts as a *direct pumping source* in surface 4WM.

In the context of nonlinear 4WM, the proposed concept is well-grounded: Now that the emitted 4WM field can be purely evanescent (nonradiative) as confirmed in Refs. [17-19], so can be any one of the 4WM pumping fields. It is therefore conceivable that the SPPs-based 4WM processes should be able to effectively take place at the dielectric/metal interface, provided that all the involved fields are coherent, *irrespective of being optical or evanescent*. It can also be expected to take full advantage of the local-field enhancement effect induced by the plasmon resonance of the metal on smooth surfaces to considerably enhance its conversion efficiency, a feat that would otherwise be unattainable in conventional optical 4WM processes.

Here we should mention that similar processes using SPPs to serve as the input waves have already been experimentally demonstrated in the coupling of optical SHG to SPPs in thin silver films [10], the surface CARS of the benzene/silver system [14], and the surface 4WM at the air/gold interface using only incident optical fields as the pumping sources [18,19], all of which feature SPP fields by themselves acting as input waves, without extra optical fields interacting with them, to yield the desirable enhanced output of SHG, CARS, or 4WM. However, the concept developed here features optically launched SPP field acting as a direct pumping source that coherently interacts with an extra propagating optical field to generate another outgoing optical field (the emitted 4WM radiation in this particular case). It represents one of the possible combinations where one of the two interacting fields is evanescent and the other one is propagating in a dielectric; nonetheless, the case investigated here has not been specifically analyzed so far.

**2. Local-field enhancement effect**

We consider here the SPPs-based 4WM process that uses a light-excited SPP field and an incident optical field as the input waves at the air/gold film interface. Of particular interest are the two schemes [(a) and (b)] depicted in the top panels of Fig. 1:

$$\omega_{4\text{wm(a)}} = 2\omega_{\text{spp}} - \omega_{\text{opt}} \quad [\text{scheme (a)}], \tag{1}$$

and
$$\omega_{4\text{wm(b)}} = 2\omega_{\text{opt}} - \omega_{\text{spp}} \quad [\text{scheme (b)}]. \tag{2}$$

Here, the SPP field with frequency $\omega_{\text{spp}}$ is launched by an external optical field ($\lambda_{\text{exc}}$) in the well-known Kretschmann-Raether (KR) geometry [22]. A second optical field at $\omega_{\text{opt}}$ is introduced to interact with the SPP field at the air/gold film interface, giving rise to the coherent 4WM radiative fields with frequencies $\omega_{4\text{wm(a)}}$ or $\omega_{4\text{wm(b)}}$. The third-order susceptibility of metallic gold, $\chi_{\text{Au}}^{(3)}$, induces nonlinear surface polarization (expressed in its scalar form [19]) for each scheme as follows:

$$P^{(a)}(\omega_{4wm(a)} = 2\omega_{spp} - \omega_{opt}) = \varepsilon_0 \chi_{Au}^{(3)}(-\omega_{4wm(a)}; \omega_{spp}, \omega_{spp}, -\omega_{opt}) \times$$
$$E_{spp}(\omega_{spp})E_{spp}(\omega_{spp})E_{opt}^*(\omega_{opt}) \quad (3)$$

and
$$P^{(b)}(\omega_{4wm(b)} = 2\omega_{opt} - \omega_{spp}) = \varepsilon_0 \chi_{Au}^{(3)}(-\omega_{4wm(b)}; \omega_{opt}, \omega_{opt}, -\omega_{spp}) \times$$
$$E_{opt}(\omega_{opt})E_{opt}(\omega_{opt})E_{spp}^*(\omega_{spp}) \quad (4)$$

where $E_{spp}(\omega_{spp})$ and $E_{opt}(\omega_{opt})$ are the electric fields associated with the in-plane SPP wave and the propagating $\omega_{opt}$ beam, respectively, and $\varepsilon_0$ is the permittivity of free space.

If the SPP field in scheme (a) is replaced with the propagating optical field ($\lambda_{exc}$) that is originally used to excite the SPP wave, the corresponding 4WM process [denoted (a′)] may, alternatively, be expressed by a nonlinear polarization according to

$$P^{(a')}(\omega_{4wm(a')} = 2\omega_{exc} - \omega_{opt}) = \varepsilon_0 \chi_{Au}^{(3)}(-\omega_{4wm(a')}; \omega_{exc}, \omega_{exc}, -\omega_{opt}) \times$$
$$E_{exc}(\omega_{exc})E_{exc}(\omega_{exc})E_{opt}^*(\omega_{opt}) \quad (5)$$

where $E_{exc}(\omega_{exc})$ is the electric field associated with the SPP-excitation light ($\lambda_{exc}$) that carries exactly the same frequency as the SPP wave does, a consequence of energy conservation, i.e., $\omega_{exc}$ (= $2\pi c/\lambda_{exc}$) ≡ $\omega_{spp}$. [Note here, however, that $\lambda_{spp} \neq \lambda_{exc}$; cf. Eqs. (9) and (12) given below.] Eq. (3) differs from Eq. (5) only in the two $E_{spp}(\omega_{spp})$ fields, and hence the intensity enhancement, represented by an amplification factor $G$, under this particular scheme reads

$$G = \left|\frac{E_{spp}(\omega_{spp})}{E_{exc}(\omega_{spp})}\right|^4, \quad (6)$$

which may be estimated to be $10^4$–$10^8$ as the SPP field, $E_{spp}(\omega_{spp})$, is usually 1–2 orders of magnitude stronger than the electric field of the SPP-excitation light, $E_{exc}(\omega_{exc})$ [3]. It is generally agreed that the physics underlying such high enhancements stems from the excitation of electromagnetic resonances in the metallic structures that permits a highly concentrated field at the very surface of metal film [3,4]. [Alternatively, if switching from scheme (a) to scheme (b) one would still expect a quite high enhancement (to the second power of $|E_{spp}(\omega_{spp})/E_{exc}(\omega_{spp})|$).] As Novotny and coworkers well investigated both theoretically and experimentally, the 4WM field sustains an evident enhancement where the highest conversion efficiency is found when the evanescent 4WM wave couples to SPPs [19]. Moreover, it is well known that enhancement can be experienced when SPP waves are closer to the plasmon resonance of the metal.

It is worth pointing out that the local-field enhancement effect for the two schemes considered here may not be as predominant as that for other third-order nonlinear processes with three SPP fields serving as the input waves (e.g., the case in surface CARS [14]). Nonetheless, we should stress here that the main theme of this contribution lies in exploring the well-groundedness and feasibility of the developed concept, not focusing only on the local-field enhancement effect.

## 3. Conceptual experiments at the air/gold film interface

Equations (1) and (2) are the statements of energy conservation, defining the frequencies of the outgoing 4WM radiation. To effectively generate the 4WM emission, one must also ensure momentum conservation of the four fields involved. In the very recent works [18-20] where two incident laser beams with frequencies $\omega_1$ and $\omega_2$ give rise to 4WM fields with frequencies $2\omega_1 - \omega_2$ and $2\omega_2 - \omega_1$ on gold surfaces, it is sufficient to take into account only the momentum conservation along the interface because it is the evanescent energy of the in-plane SPP wave induced by the nonlinear response of gold that converts via nanostructured surfaces into the propagating 4WM radiation. In our case, however, the outgoing 4WM emission results from mixing two SPP fields with one optical field [scheme (a)] or mixing one SPP field with two optical fields [scheme (b)]. The resulting 4WM radiation discussed here, unlike the in-plane evanescent 4WM wave coupled to SPPs as investigated in Ref. [19], is generated via direct nonlinear interactions of the SPP and optical fields. Therefore, the momentum conservation conditions for our case are essentially different, that is, in order for the proposed concept to work effectively, not only the in-plane but also the out-of-plane momenta (of incident photons and/or in-plane SPPs) must be strictly conserved [cf. Eqs. (7) and (8) given below, which relate to the parallel in-plane wave vector and the orthogonal vertical component, respectively].

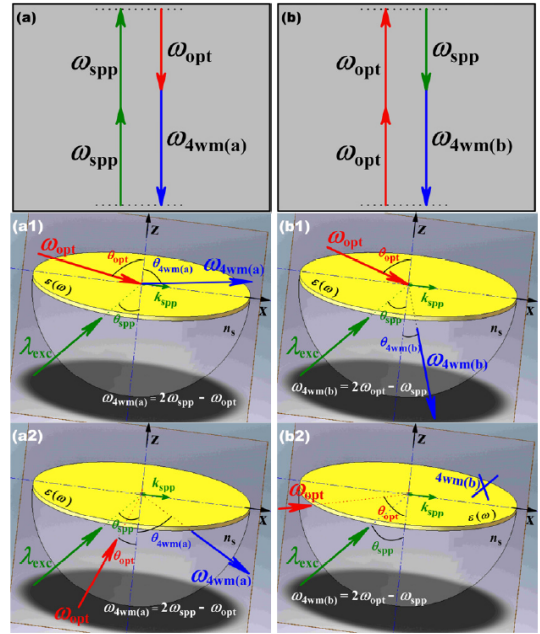

Fig. 1. Diagram of the two SPPs-based 4WM schemes and sketches of the corresponding configurations for the case of air/gold film interface (see text for details).

Let us first consider scheme (a). Depending on the incidence directions of the $\omega_{opt}$ beam two possible configurations may be considered, as sketched in panels (a1) and (a2) of Fig. 1. Propagation of the $\omega_{opt}$ beam is kept in-plane with that of the SPP-excitation beam ($\lambda_{exc}$), and both beams are chosen to be *p*-polarized. [The *s*-polarized $\lambda_{exc}$ beam can hardly excite SPPs, while the *s*-polarized $\omega_{opt}$ beam will dramatically reduce the 4WM conversion efficiencies (cf. Fig. 3 given below).] Without losing generality, the $\lambda_{exc}$ beam is confined in the third quadrant of x-z plane perpendicular to the metal surface, making the SPP wave vector pointing to the +x direction. All the angles are measured from the surface normal in clockwise direction. To excite SPPs at the air/gold film interface, we here use a modified KR configuration that consists of a hemisphere glass prism and a thin film of gold (thickness typically below 100 nm) deposited on the planar surface of the glass substrate. The hemisphere geometry is adopted so as to facilitate discrepancy-free angular measurements, which is crucial when optical beams traveling into or out of the glass substrate are dealt with. It is important to note that the two laser beams must be precisely synchronized to within the SPP launch window as short as 1 ps [23]. However, depending on the time duration of the laser pulses, 1-ps synchronization may not be sufficient to create optimal 4WM signals, especially when using femtosecond pulses. Importantly also, considering that SPP propagation lengths along the metal surface are generally restricted to the order of 10 μm by the high SPP damping due to ohmic losses in the metal [3], one should also ensure a well-defined spatial overlap of the two beams focused at the very center of the metal surface. Technically, all these are not difficult to accomplish.

As for the first configuration [Fig. 1(a1)] where the $\omega_{opt}$ beam impinges on the metal surface from the air side, momentum conservation defines the propagation directions according to

$$k_{4wm(a)}^{//} = k_{4wm(a)} \sin\theta_{4wm(a)} = 2\operatorname{Re}(k_{spp}) + k_{opt} \sin\theta_{opt}, \quad (7)$$

$$k_{4wm(a)}^{\perp} = k_{4wm(a)} \cos\theta_{4wm(a)} = k_{opt} \cos\theta_{opt}, \quad (8)$$

where $\quad k_{4wm(a)} = \dfrac{\omega_{4wm(a)}}{c}, \ k_{opt} = \dfrac{\omega_{opt}}{c}, \ \text{and} \ k_{spp} = \dfrac{\omega_{spp}}{c}\sqrt{\dfrac{\varepsilon_d \varepsilon_m(\omega_{spp})}{\varepsilon_d + \varepsilon_m(\omega_{spp})}}, \quad (9)$

$\varepsilon_m(\omega_{spp})$ is the frequency-dependent complex dielectric function of the metal (data for gold taken from Ref. [24]) and $\varepsilon_d$ is the dielectric constant of the superstrate material ($\varepsilon_d = 1$ for air). Equations (7)–(9) combined with Eq. (1) yield the two sets of angle of interest:

$$\theta_{opt} = \arcsin\left(\frac{1-\eta^2}{\eta}\frac{\lambda_{opt}}{\lambda_{exc}} - \frac{1}{\eta}\right), \quad (10)$$

$$\theta_{4wm(a)} = \arcsin\left(\frac{1+\eta^2}{\eta}\frac{\lambda_{4wm(a)}}{\lambda_{exc}} - \frac{1}{\eta}\frac{\lambda_{4wm(a)}}{\lambda_{opt}}\right), \quad (11)$$

where $\quad \lambda_{exc} = \dfrac{2\pi c}{\omega_{spp}}, \ \lambda_{opt} = \dfrac{2\pi c}{\omega_{opt}}, \ \lambda_{4wm(a)} = \dfrac{2\pi c}{\omega_{4wm(a)}}, \quad (12)$

and $\quad \eta = \operatorname{Re}\left(\sqrt{\dfrac{\varepsilon_d \varepsilon_m(\omega_{spp})}{\varepsilon_d + \varepsilon_m(\omega_{spp})}}\right). \quad (13)$

Note that $\eta[= \eta(\lambda_{exc}) = \eta(\omega_{spp})]$, usually referred to as the effective refractive index [25], is the real part of the ratio of $k_{spp}$ to the free-space wave vector $[2\pi/\lambda_{exc} (= \omega_{spp}/c)]$. The wavelength-dependent SPP-excitation angle, $\theta_{spp}(\lambda_{exc})$, is simply determined by

$$\sin\theta_{spp}(\lambda_{exc}) = \frac{\eta(\lambda_{exc})}{n_s}, \quad (14)$$

where $n_s$ is the refractive index of the substrate ($n_s \approx 1.5$ for BK7 glass used here). This specific incidence angle allows matching of the SPP wave vector by the light wave-vector component parallel to the surface, thereby generating the evanescent SPP wave at the metal surface [1-4].

When both $\lambda_{exc}$ and $\lambda_{opt}$ are scanned from 350 to 1200 nm in a 1 nm step, Eqs. (10) and (11) yield the simulated 2D-WAD spectra of $\theta_{opt}(\lambda_{exc}, \lambda_{opt})$ [Fig. 2(a)] and $\theta_{4wm(a)}(\lambda_{exc}, \lambda_{opt})$ [Fig. 2(b)]. It is seen that the momentum conservation restricts $\lambda_{exc}$ to below 417 nm (while $\lambda_{opt}$ could be anywhere in 350–1200 nm), and hence $\lambda_{4wm(a)}$ to within a range of 205–515 nm [Fig. 2(c)]. Interestingly, in this particular configuration the propagation of the $\lambda_{opt}$ beam must be confined to the second quadrant ($\theta_{opt} < 0$) and is nearly parallel to the surface (close to the so-called "grazing incidence"), while the outgoing 4WM radiation is expected to be detectable only in the first quadrant ($\theta_{4wm(a)} > 0$) and also nearly parallel to the surface [$\theta_{4wm(a)} \in (84.3°, 89.5°)$]. Admittedly, such an observation together with the very narrow (< 6°) range available for $\theta_{4wm(a)}$ may affiliate, to a certain extent, the proof-of-principle experiments using this configuration. However, this can be easily overcome by means of narrow bandpass filtering combined with routinely available high-sensitivity detection techniques (e.g., photon counting).

Note that the *d*-band transition resonance for gold centers at ~2.38 eV (~520 nm) and hence 520 nm is usually referred to as the typical wavelength for inducing surface plasmon resonance (SPR) on gold surfaces [23]. The wavelengths below 520 nm are known to be less efficient than 520 nm (and above) for launching SPPs at the air/gold film interface; nonetheless, these wavelengths do work in principle provided that Eq. (14), the total internal reflection condition indispensable to the excitation of SPPs in the KR geometry, can be fulfilled for the spectral range of 350–520 nm. As a matter of fact, there existed reports using wavelengths in this range for the air/gold film system, e.g., 514 nm [26] and 400 nm [27], to name a few. As our paper mainly addresses conceptual aspects, providing numerical simulations in a broader wavelength range (including here 350–520 nm) can be thought of as being appropriate.

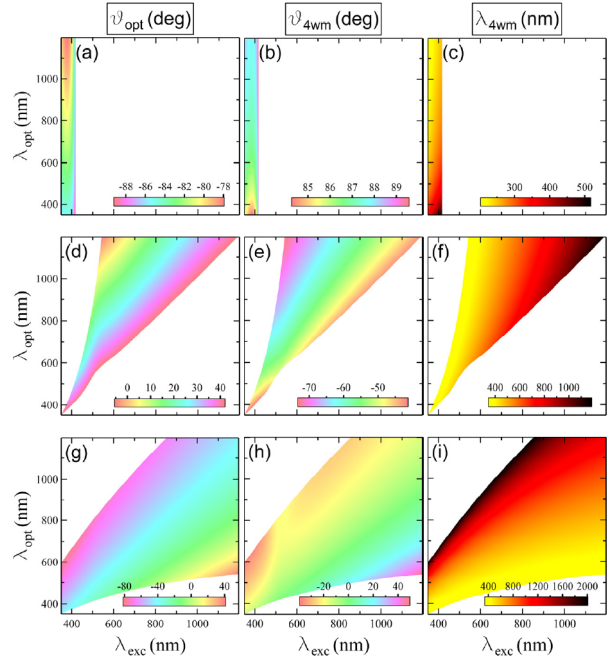

Fig. 2. 2D-WAD spectra of $\theta_{opt}(\lambda_{exc}, \lambda_{opt})$ (left column) and $\theta_{4wm}(\lambda_{exc}, \lambda_{opt})$ (middle column), and the corresponding 4WM wavelength distributions (right column) for the case of air/gold film interface. The top, middle, and bottom rows correspond, respectively, to the configurations (a1), (a2), and (b1) shown in Fig. 1.

We turn now to the second configuration of scheme (a), as depicted in Fig. 1(a2). Unlike the first one, this configuration features an $\omega_{opt}$ beam incident from the glass (instead of air) side. A simple vectorial algebra reveals immediately that the outgoing 4WM beam (if existing) has to radiate also in the lower half-space. The fact that both beams propagate in the lower half-space suggests that (i) $n_s$ must be taken into account, (ii) $\lambda_{4wm(a)} > 350$ nm due to the transmission property of BK7 glass, and (iii) those nominal solutions of ($\lambda_{exc}$, $\lambda_{opt}$) with corresponding $\theta_{opt}$ equal to or larger than the total internal reflection angles should be invalidated. Despite these constraints the obtained spectra [Figs. 2(d)–2(f)] turn out to permit a much better angle and wavelength tunability: $\theta_{opt} \in (-5.2°, 41.8°)$, $\theta_{4wm(a)} \in (-73.1°, -41.4°)$, and $\lambda_{4wm(a)} \in [350\ nm, 1182\ nm]$.

Although scheme (b) is less competitive than scheme (a) in terms of the ability to increase the SPPs-based 4WM conversion efficiency, it is still capable of bringing forth intensity enhancement that is quite pronounced. Therefore, it is necessary to gain insight into its feasibility, at least for the sake of completeness. This scheme may also be arranged in two configurations, as sketched in Figs. 1(b1) and 1(b2). Simulations indicate that although the configuration shown in Fig. 1 (b2) does not work, the one shown in Fig. 1(b1) not only works but offers a best tunability: $\theta_{opt} \in (-81.2°, 40.6°)$, $\theta_{4wm(b)} \in (-38.9°, 49.2°)$, and $\lambda_{4wm(b)} \in [350\ nm, 2000\ nm]$ [Figs. 2(g)–2(i)].

Interestingly, among the four configurations of the two schemes discussed above, the only one that fails to generate the 4WM radiation belongs to scheme (b). As for the preferred scheme (a), from an experimental point of view, it would be highly recommended to adopt the second configuration in light of the better accessibility and tunability for both wavelength and angle it offers.

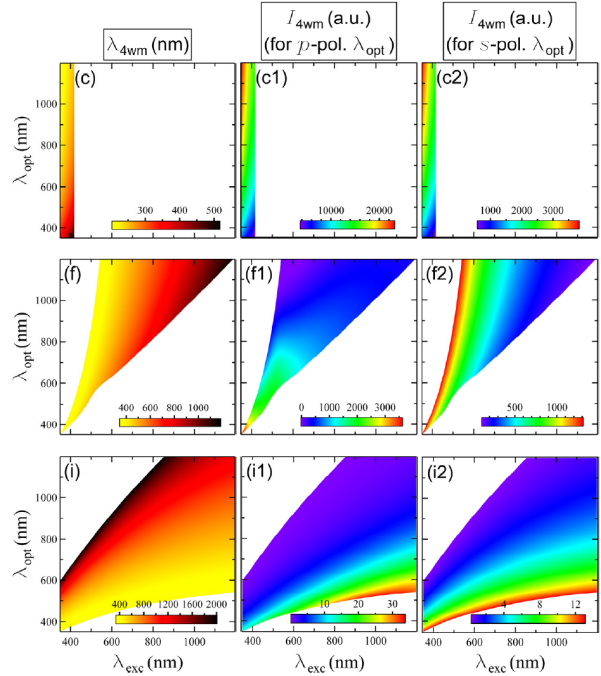

Fig. 3. The 4WM wavelength distributions [(c), (f), and (i), see also Fig. 2] and their corresponding 4WM conversion efficiencies for two particular cases at the air/gold film interface: (1) the $\lambda_{opt}$ beam is $p$-polarized [(c1), (f1), and (i1)]; (2) the $\lambda_{opt}$ beam is $s$-polarized [(c2), (f2), and (i2)]. As in Fig. 2, the top, middle, and bottom rows correspond, respectively, to the configurations (a1), (a2), and (b1) as sketched in Fig. 1. The scalar bar in each panel of the middle and right columns shows the relative 4WM intensities in arbitrary units.

We have further performed numerical simulations for the conversion efficiencies (i.e., the intensity distributions) of the three 4WM processes (whose 2D-WAD spectra and 4WM wavelength distributions are shown in Fig. 2). The simulated results (under a conservative estimation of $|E_{spp}| = 10 \times |E_{opt}|$ [3], as an example) are illustrated in the middle column (for $p$-polarized $\lambda_{opt}$ beam) and the right column (for $s$-polarized $\lambda_{opt}$ beam) of Fig. 3. [The left column just replots the corresponding 4WM wavelength distributions for convenient tracking.]

These intensity distributions are obtained following a careful analysis with respect to the rich structure (81 components) of $\chi_{Au}^{(3)}$ (a tensor of rank four), detailed description of which (not given here) can be found in Ref. [19]. Not surprisingly, the cases with a *p*-polarized $\lambda_{opt}$ beam [Figs. 3(c1), 3(f1), and 3(i1)] possess higher 4WM conversion efficiencies than those with an *s*-polarized $\lambda_{opt}$ beam [Figs. 3(c2), 3(f2), and 3(i2)]. Also, as expected, the cases using scheme (a) [the two upper rows of Fig. 3] are found to bear much higher efficiencies than those using scheme (b) [the bottom row of Fig. 3]. The intensity simulations may provide another set of informative guiding maps, in addition to the 2D-WAD spectra shown in Fig. 2, for further experimental verification of the concept developed in this paper.

### 4. An extension to the graphene flake/gold film interface

The conceptual experiments presented in this paper are associated with, but not limited to, the air/gold film interface. On the one hand, since the proposed concept is based on SPPs, it may also apply to other metallic materials than gold (e.g., silver and copper). Our choice of gold here is mainly based on its several merits: (i) $\chi_{Au}^{(3)}$ is quite large [28,29], (ii) its chemical stability is high and its handling is convenient [4], and (iii) the resonant SPP-excitation on a gold film is usually in the desirable visible (or near infrared) spectral region [3]. On the other hand, when air ($\varepsilon_{air} = 1$) is replaced with other dielectric materials ($\varepsilon_d$), the concept may also be feasible given that SPPs can be launched and sustained by such dielectric/metal surfaces. In the following we present an extended case study of SPPs-based 4WM at the graphene flake/thin gold film interface.

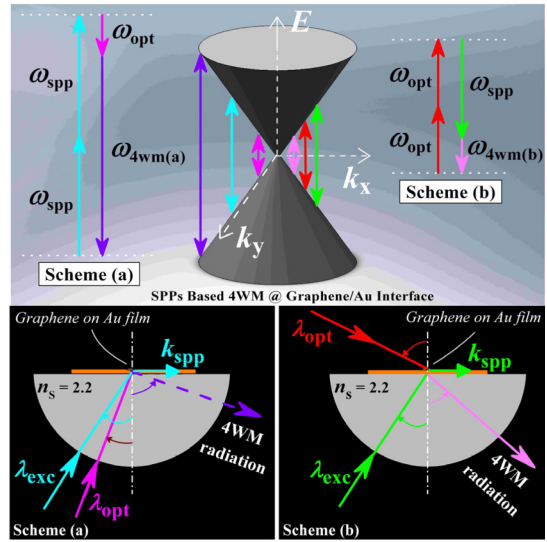

Fig. 4. Upper panel: Schematic of graphene's band structure with the three resonant-field energies (double arrows) involved in the two schemes of SPPs-based 4WM. Lower panels: Sketches of the two working configurations for scheme (a) (left) and scheme (b) (right) when applied to the graphene flake/thin gold film system.

Being a strictly 2D material, graphene offers inroads into a wealth of new physics [30-32]. In a very recent investigation on the nonlinear optical properties of graphene by means of 4WM, the effective $\chi_{Gr}^{(3)}$ of graphene flakes has been determined to be ~8 orders of magnitude and ~25-fold larger than that of the insulating glass material and the quasi-2D thin film of gold, respectively [33]. We thus envisage that such an exceptionally high nonlinear response of graphene would allow for another 1–2 orders of magnitude enhancement of the third-order nonlinear conversion efficiency, provided that the marriage of our concept and the graphene flake/gold film interface is feasible. More importantly, because interband electron transitions in graphene can occur at all energies as a result of its unique linear band structure [30-32], such a marriage (if proved successful) can be expected to generate both frequency and angle tunable, significantly enhanced, nonlinear surface 4WM radiation, hence showing enticing potentials for graphene-based nanophotonics and nanoplasmonics.

As in the air/gold film case, SPPs-based 4WM at the graphene flake/gold film interface may also be implemented through two schemes, as depicted in the top panel of Fig. 4. Since the SPP field along the +z direction retains within the nanometer scale (the penetration length of the SPP field in the adjacent media is in the order of the SPP wavelength or larger) [3], the graphene sample deposited on the gold film can be multi-layered flakes. It is worth noting that the refractive indexes of graphene flake ($n_{Gr}$) in the 300–1000 nm spectral region have recently been determined to be larger than 2.6 [34], which inevitably necessitates a substrate material with a rather large $n_s$ value so as to effectively excite SPPs at the graphene flake/gold film interface [cf. Eq. (14)]. For our proof-of-principle conceptual experiments we here choose cubic zirconia ($ZrO_2$) whose $n_s$ values are ~2.2 in a wavelength range of 400–525 nm where light transmission is extremely high (~100%) [35]. Note also that over the entire 125-nm-wide wavelength range $n_{Gr}$ of the multi-layered graphene flake takes a nearly constant value of ~2.7 (cf. Fig. 2 of Ref. [34]), which corresponds to a dielectric constant of $\varepsilon_{Gr} \approx 7.3$. The use of this nearly constant $\varepsilon_{Gr}$ value obviously facilitates our numerical simulations.

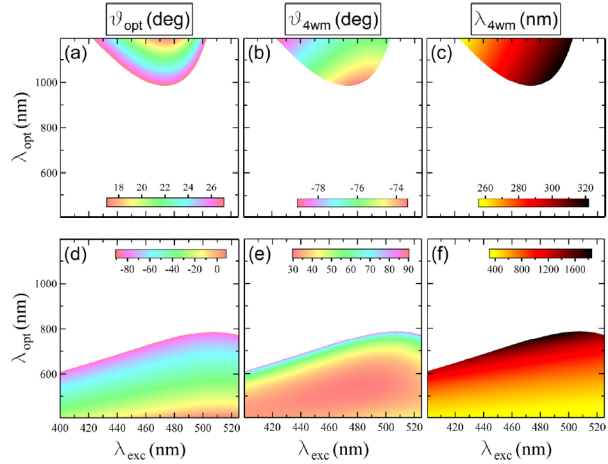

Fig. 5. 2D-WAD spectra of $\theta_{opt}(\lambda_{exc}, \lambda_{opt})$ (left column) and $\theta_{4wm}(\lambda_{exc}, \lambda_{opt})$ (middle column), and the corresponding 4WM wavelength distributions (right column) for the case of graphene flake/gold film interface. The upper and lower rows correspond, respectively, to schemes (a) and (b) shown in Fig. 4.

Our simulations indicate that scheme (a) works solely for the configuration in which the $\omega_{opt}$ beam is incident from the lower half-space, as depicted in the bottom-left panel of Fig. 4. The obtained spectra are shown in Figs. 5(a)–5(c), from which one can readily detect that the angle tunability is quite limited [$\theta_{opt} \in (17°, 27°)$ and $\theta_{4wm(a)} \in (-79°, -73.4°)$]; nonetheless, a fairly broad (~65-nm) tunable range of $\lambda_{4wm(a)}$ can be achieved when $\lambda_{exc}$ and $\lambda_{opt}$ are scanned in 400–525 nm and 400–1200 nm, respectively. Unfortunately, the resulting 4WM radiation that also propagates in the lower half-space is generated in the ultraviolet (256–321 nm) where $ZrO_2$ significantly absorbs the light [35]. [As such, the outgoing 4WM emission depicted in the bottom-left panel of Fig. 4 is indicated by a dashed line.] One solution to this problem is simply to replace $ZrO_2$ with UV-grade fused silica ($SiO_2$) for this half portion of hemisphere. Alternatively, to make this configuration be of practical value, one may resort to other high-$n_s$ materials with good optical transmission for the emitted 4WM wavelengths, namely, the lack of practicality in this specific (unlucky) case will not harm the validity of scheme (a) for the considered graphene flake/gold film system.

We emphasize here again that thus obtained 4WM radiation is estimated to be enhanced by 4–8 orders of magnitude (due to the local-field enhancement effect) when compared to that achieved at the graphene flake/glass interface [33], and by 1–2 orders of magnitude when compared to that achieved at the air/gold film interface (due to $\chi_{Gr}^{(3)}/\chi_{Au}^{(3)} \approx 25$ [29,33]). It is worth mentioning here that Novotny and co-workers have very recently demonstrated that, by depositing a thin dielectric layer on the metal surface or, alternatively, using a thin metal film, one can enhance 4WM in metals by up to 4 orders of magnitude [20].

Scheme (b), on the contrary, turns out to work only for the configuration where the $\omega_{opt}$ beam is incident from the upper half-space, as depicted in the bottom-right panel of Fig. 4. The obtained spectra are shown in Figs. 5(d)–5(f). Notwithstanding scheme (b) is less efficient than scheme (a) in terms of the enhancement $G$ factor, it not only compensates for the lack of angle tunability [$\theta_{opt} \in$ (-89.9°, 7.5°) and $\theta_{4wm(b)} \in$ (30°, 89.8°)] but enables a much better wavelength tunability ($\lambda_{4wm(b)} \in$ [323 nm, 1845 nm]). The two schemes appear to be complementary to each other when applied to this graphene flake/gold film interfacial system.

Last but not least, obviously the concept developed here can be readily extended to other dielectric/metal interface and to other surface nonlinear processes, e.g., sum-frequency generation (SFG). As a second-order nonlinear process, SFG is dependent on the second-order susceptibility $\chi^{(2)}$ which becomes zero in free-standing graphene as it is a strictly centrosymmetric 2D medium. However, at an interface between graphene flake and gold film, the inversion symmetry will be broken and an SFG signal can be created. Interestingly, if such a graphene flake/gold film interfacial system is combined with the SPPs-based concept developed here, one would expect to achieve an exceptionally high conversion efficiency for surface SFG, much higher than that for surface 4WM described in this paper as $\chi^{(2)} >> \chi^{(3)}$.

## 5. Concluding remarks

In summary, we have developed a concept of surface plasmon polaritons (SPPs) based four-wave mixing (4WM). The working mechanism of this concept distinguishes itself in that, as a direct pumping source (evanescent yet coherent) for the third-order nonlinear process, the nonradiative SPP field coherently interacts with a radiative optical field at the dielectric/metal interface, imparting local-field enhancement effect to the resulting 4WM field thereby dramatically increasing the surface 4WM conversion efficiencies. Conceptual experiments have been performed through numerical simulations for various possible configurations at the air/gold film interface with their feasibility carefully examined. We have further exploited the possibilities of incorporating the concept with the unique linear band structure as well as remarkably strong nonlinear optical response of graphene, towards generating significantly enhanced, both frequency and angle tunable, directional, and coherent surface 4WM radiation. We believe and hope that the proposed concept (and its effective integration with graphene flakes or other nanomaterials as well as its possible extension to other nonlinear processes) may open opportunities for applications in newly emerging nanophotonics, optoelectronics, and active plasmonics.

**Acknowledgments**

This work is supported by the National Basic Research Program of China (2010CB923300, 2007CB815203), the National Natural Science Foundation of China (20873133, 20925311), the Chinese Academy of Sciences (KJCX2-YW-N24), and the Fundamental Research Funds for the Central Universities of China (WK2340000012). The authors wish to thank X.-P. Wang for stimulating discussions and the anonymous reviewers for helpful suggestions.